\documentclass[aps,pra,longbibliography,groupedaddress]{revtex4-1}

\usepackage{graphicx}
\usepackage{epstopdf, epsfig}
\usepackage{natbib}            		
\usepackage{color}
\usepackage[%
colorlinks=true,
urlcolor=blue,
linkcolor=blue,
citecolor=blue
]{hyperref}
\usepackage[usenames,dvipsnames,svgnames,table]{xcolor}
\usepackage{booktabs}	
\usepackage{amsmath}	
\usepackage{bm}
\usepackage{siunitx}	
\usepackage{tikz}		
\usetikzlibrary{patterns}
\usetikzlibrary{arrows}
\usetikzlibrary{decorations.markings}
\tikzstyle streamline1=[postaction={decorate,decoration={markings,
		mark=at position .5 with {\arrow{stealth}}}}]
\tikzstyle streamline2=[postaction={decorate,decoration={markings,
		mark=at position .66 with {\arrow{stealth}},mark=at position .33 with {\arrow{stealth}}}}]

\begin{document}


\title{Entrainment effects in periodic forcing of the flow over a backward-facing step}


\author{T. Berk}
\author{T. Medjnoun}
\author{B. Ganapathisubramani}
\email{G.Bharath@southampton.ac.uk}
\affiliation{Aerodynamics and Flight Mechanics group, University of Southampton}


\date{\today}

\begin{abstract}
The effect of the Strouhal number on periodic forcing of the flow over a backward-facing step  (height, $H$) is investigated experimentally.
Forcing is applied by a synthetic jet at the edge of the step at Strouhal numbers ranging from $0.21<St_H<1.98$ ($St_H = f H/U_\infty$) at a Reynolds number of $Re_H = HU_\infty/\nu = 41000$. 
In the literature, the effect of Strouhal number on the reattachment length is often divided into low- and high frequency actuation, referring to different frequency modes in the unforced flow.
In the present paper, variations with Strouhal number are explained based on entrainment rather than frequency modes.
The reattachment length is shown to decrease linearly with entrainment.
Entrainment is driven by vortices generated by the forcing and locally entrainment is shown to be qualitatively similar to circulation for all cases considered.
Total circulation (and therewith entrainment and the effect on the reattachment length) is shown to decrease with Strouhal number whereas this is not predicted by models based on frequency modes.
An empirical model for the (decay of) circulation is derived by tracking vortices in phase-locked data. 
This model is used to decipher relevant scaling parameters that explain the variations in circulation, entrainment and reattachment length.
Three regimes of Strouhal number are identified.
A low-Strouhal-number regime is observed for which vortices are formed at a late stage relative to the recirculation region, causing a decrease in effectiveness.
For high Strouhal numbers vortices are being re-ingested into the actuator or are packed so close together that they cancel each other, both decreasing the effectiveness of forcing.
In the intermediate regime a vortex train is formed of which the decay of circulation increases for increasing Strouhal number.
The scaling of this decay fully explains the observed variation in reattachment length.
The observations on entrainment made in this study are expected to also hold for periodic forcing of other bluff-body flows. 
\end{abstract}


\maketitle

\section{Introduction\label{sec:Intro}}
The flow over a backward-facing step separates from the edge of the step and reattaches further downstream.
Aerodynamic drag as caused by separation can account for more than 50\% of fuel consumption in transport~\citep{Hucho1993}.
Most drag reduction strategies focus either on increasing the pressure recovery or on streamlining the body using elongation of the wake~\citep{Barros2016a}.
The present paper focusses on the latter by studying the effect of forcing on the length of the wake.

The size of the wake is quantified by defining a recirculation region where the local (mean) velocity is in the direction opposite to the mean flow ($\widetilde{U}_x < 0$).
The reattachment point is defined as the point on the wall where the time-averaged horizontal flow switches back from negative to positive (i.e. $\widetilde{U}_x = 0$).
The length of the recirculation region (or reattachment length) is the distance from the base of the step to the reattachment point.

This length can be manipulated using local forcing of the flow.
Examples of forcing methods applied in the literature are an oscillating flap \citep{Nagib1985}, steady blowing \citep{Littlewood2012}, steady suction \citep{Sano2009}, pulsed jets \citep{Barros2016a,Barros2016}, pulsed plasma actuators \citep{Roupassov2009} or synthetic jets \citep{Chun1996,Dandois2007,Henning2007,Dahan2012,Oxlade2015}.
Apart from the steady suction and blowing, these are all periodic forcing methods.
Particularly for high forcing frequencies \citep{Dandois2007,Vukasinovic2010}, this periodic forcing has the ability to dampen the flow unsteadiness and stabilise the shear layer, which is assumed to lower entrainment and elongate the recirculation region.
However, periodic forcing typically generates a train of vortices, leading to an increase in entrainment and shortening the recirculation region.
The balance between these two effects determines whether a net increase or decrease in reattachment length is achieved.

In recent years, feedback and feed-forward control strategies have been developed to selectively control disturbances \citep{Herve2012,Gautier2014}.
It has been shown numerically that the turbulent kinetic energy of the incoming flow can be reduced by more than 90\% using feed-forward control \citep{Herve2012}.
A reduction of the amplitude of incoming perturbations has been shown to lead to a decrease in entrainment, which increases the reattachment length.
Machine learning control was used experimentally by \citet{Gautier2015} to optimize the feedback control law for the flow over a backward-facing step.
This optimized control performs similar to periodic forcing at design conditions and outperforms periodic forcing for off-design conditions.
Their experiments show the technical feasibility of real-time control at a Reynolds number of $Re_H = 1350$.
The present research focusses on harmonic forcing at higher Reynolds numbers which are closer to Reynolds numbers associated with practical applications.

Periodic forcing of the flow over a backward-facing step using a spoiler-like flap was investigated by \citet{Nagib1985}.
The reattachment length showed to scale with the Strouhal number, defined as

\begin{equation}
St_H = \frac{f H}{U_\infty}, \label{eq:StH}
\end{equation}

\noindent where $f$ is the forcing frequency, $H$ is the step height and $U_\infty$ is the free-stream velocity.
The same scaling is used by \citet{Kiya1993}, who investigated forcing of the leading-edge separation of a blunt axisymmetric body using synthetic jets.
They found that (for a constant Strouhal number) an increase in free-stream Reynolds number or in forcing amplitude increases the effect of forcing.

\begin{figure}
\includegraphics{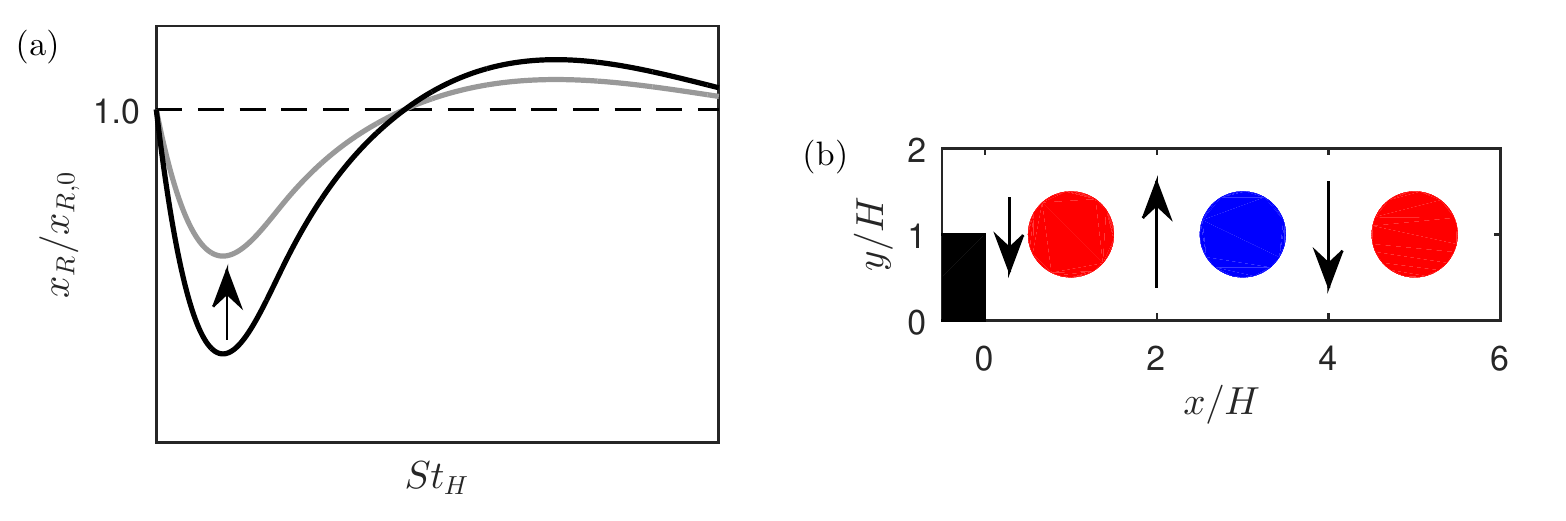}
\caption[Schematic of parameter effects on reattachment length]{
		(a) Schematic of the effect of Strouhal number on reattachment length.
		The dashed line indicates the unforced reattachment length.
		The arrow indicates a decrease in either the free-stream Reynolds number or the actuation amplitude.
		(b) Schematic of alternating positive and negative vortices with corresponding induced velocities between them as leading mechanism for entrainment.
		\label{fig:St_Xr_schem}
	}	
\end{figure}

Similar results are obtained by \citet{Chun1996} for forcing of the flow over a backward-facing step using synthetic jets.
For a range of free-stream Reynolds numbers the reattachment length, $x_R$, decreases sharply for low Strouhal numbers, reaching a minimum around ${St_H = 0.2}$--$0.3$ where the reattachment length is decreased to as little as $x_R/x_{R,0} = 0.65$, where $x_{R,0}$ represents the unforced length.
Beyond this minimum, for increasing Strouhal numbers, the reattachment length recovers, reaching the initial value around ${St_H = 0.8}$--$1.0$ and overshooting slightly to as much as $x_R/x_{R,0} = 1.05$.
This is presented schematically in Figure~\ref{fig:St_Xr_schem}a, where the black curve shows the trend for a baseline case and the grey line represents a decrease in either the free-stream Reynolds number or the forcing amplitude.

In more recent years, development of both computational (DNS, LES) and experimental (PIV) techniques have made it easier to quantify velocity fields, aiding in the investigation of the exact mechanisms causing the changes in reattachment length discussed above.
\Citet{Dandois2007} used DNS and LES to investigate the frequency effect of a synthetic jet on the separation of a rounded ramp.
They distinguish two cases, labelled a low frequency (${St_H = 0.14}$) and a high frequency ($St_H = 1.15$) case.
Their data shows that for the low frequency case large spanwise vortices are formed with the flow between each vortex being reattached.
In the time-averaged sense, this leads to a reduction of the reattachment length to $x_R/x_{R,0} = 0.46$, attributed to an increase in entrainment of high-momentum fluid into the recirculation region caused by the created vortices.
In contrast, the high-frequency case does not produce clear vortices and the synthetic jet actuator is said to operate in acoustic dominated mode, decreasing the amplitude of incoming perturbations in the flow.
The reduction of perturbations stabilizes the shear layer between the cross-flow and the recirculation region and reduces entrainment, leading to an increase in reattachment length to $x_R/x_{R,0} = 1.43$.

\citet{Dahan2012} explored numerically open- and closed-loop control of a backward-facing step using synthetic jets.
The open-loop actuation shows very similar results to the experimental study by \citet{Chun1996}, decreasing the reattachment length for $St_H < 0.8$ and slightly increasing it for $0.8 < St_H < 2$.

Forcing of the wake behind a bluff axisymmetric body is studied experimentally by \citet{Oxlade2015}. The range of Strouhal numbers tested is approximately $1.0 < St_H < 6.0$ ($St_H$ is based on radius of the body).
As expected from the above mentioned literature, no significant reduction of the recirculation region is expected for this range of Strouhal numbers.
The shortest reattachment length reported is $x_R/x_{R,0} = 0.97$, showing analogy between this three-dimensional wake and a two-dimensional step.

The trend of the reattachment length as function of the Strouhal number as presented schematically in Figure~\ref{fig:St_Xr_schem}a reappears in all cases discussed above, regardless of the type of body causing the separation or the forcing method.
As discussed above, the reattachment length decreases sharply for low Strouhal numbers.
In the literature, this is attributed to a high rate of entrainment caused by large vortices \citep{Chun1996,Dandois2007}.
For increasing Strouhal numbers -- after reaching a minimum -- the reattachment length starts recovering towards the initial value.
This decrease in effectiveness is attributed to the size of vortices decreasing, leading to lower entrainment, for an increasing Strouhal number.
At some point (${St_H = 0.8}$--$1.0$, \citep{Chun1996}) the reattachment region reaches its unforced length and even overshoots slightly.
This increase in reattachment length compared to the unforced case is generally attributed to a stabilisation of the shear layer, reducing entrainment \citep{Dandois2007,Oxlade2015}.

The entrainment caused by vortices is represented schematically in Figure~\ref{fig:St_Xr_schem}b.
Alternating clockwise and counterclockwise vortices induce alternating positive and negative velocity components in the wall-normal direction.
A negative (downwards) velocity component entrains high-momentum fluid into the recirculation region, whereas a positive (upward) component extracts low-momentum fluid from the recirculation region, leading to an increase in average momentum in the recirculation region.

In the literature it is generally assumed that the creation of larger vortices (lower forcing frequency) leads to a higher rate of entrainment.
Theoretically, the circulation $\Gamma_j$~(m$^2$s$^{-1}$) created by a synthetic jet per unit time is given by

\begin{equation}
\frac{\mathrm{d}\Gamma_j}{\mathrm{d}t} = \frac{1}{2}u_j^2(t), \label{eq:dGdt}
\end{equation}

\noindent where $u_j(t)$ (m/s) is the jet velocity~\citep{Shariff1992}.
This jet velocity is assumed to be constant over the slot and harmonic in time, i.e.

\begin{equation}
u_j(t) = u \: \mathrm{sin}\left(2 \pi t f\right), \label{eq:ut}
\end{equation}

\noindent where $u$~(m/s) is the maximum jet velocity and $f$~(Hz) is the actuation frequency.
Integrating Equation~\ref{eq:dGdt} over a single blowing period leads to a circulation per velocity cycle of

\begin{equation}
\Gamma_j = \int_0^T \frac{1}{2}u_j^2(t) \mathrm{d}t = \frac{u^2}{4f}, \label{eq:Gamma}
\end{equation}

\noindent where $T = 1/f$ (s) is the period of the velocity cycle.
Since each velocity cycle creates a vortex pair \citep{Smith1998}, the number of vortices created per unit time is given by $N = 2 f$.
This means that the total circulation created per unit time scales as $\Gamma_j N \propto u^2$, which shows that although the circulation per vortex scales with the actuation frequency, the total circulation per unit time is independent of the actuation frequency.
Therefore, the assumption often made in the literature that entrainment decreases for an increasing frequency (smaller vortices) is not necessarily true.

The purpose of the present paper is to reassess the influence of Strouhal number on the reattachment length and identify the driving physical mechanisms of variations with Strouhal number.
This is studied experimentally by forcing the flow over a backward-facing step using synthetic jets, actuated at four different Strouhal numbers.

First, time-averaged velocity fields showing the reattachment region are presented and the variation of reattachment length with Strouhal number is discussed globally.
This is followed by phase-locked velocity and vorticity fields, showing a train of alternating vortices as driving mechanism for entrainment.
The formation and evolution of this train of alternating vortices is studied using phase-locked vorticity maps.
Next, the time-averaged circulation and entrainment as a result of this vortex train are presented and compared between the different cases.
Observed variations in entrainment are explained using tracked trajectories and circulation of the vortices.
Scaling factors for the development of the vortex location and circulation in time and space are derived and the trajectories and circulation are fitted against analytical and empirical formulae. Finally, a reconsideration of Strouhal-number regimes that highlights the entrainment effects is discussed.

\section{Experimental set-up and procedures}
\begin{figure}
	\begin{tikzpicture}
		\node [inner sep=0pt,above right] at (0,0)
			{\includegraphics{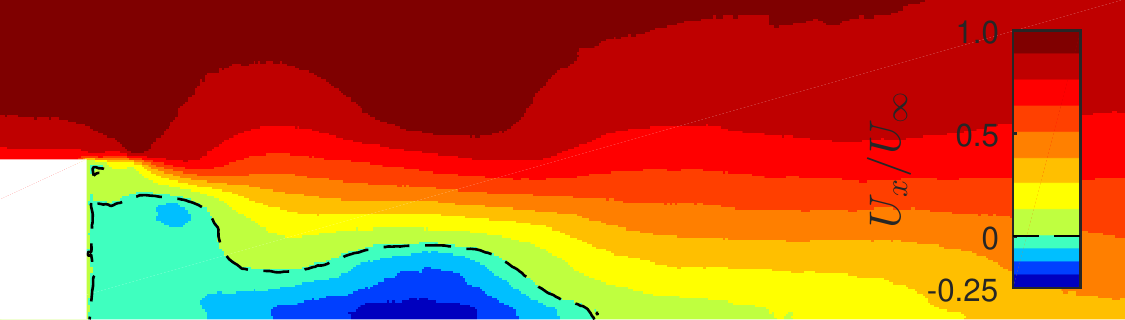}};
		\fill[white] (-.1,-.1) rectangle (0.89,1.64);
		\draw[ultra thick,black] (12,0) -- (0.89,0);
		\draw[ultra thick] (.1,1.44) --++ (-0.3,-0.45) rectangle ++ (-0.2,-0.54) ++ (0.2,0) --++ (0.3,-0.45) -- (0.89,0) (.1,0) --++ (0,1.44);
		\draw[very thick] (-0.3,0.45) to [out=270,in=0] (-1,0.6) ++ (0,-.45) rectangle ++ (-1.4,.9);
		\node at (-1.7,0.6) [align = center] {Control: \\ $f$, $u$};
		\draw[black,ultra thick, fill=lightgray] (-2,1.64) -- (0.64,1.64) -- (0.44,1.44) -- (-2,1.44);
		\draw[ultra thick, fill=lightgray] (0.89,0) -- (0.89,1.64) -- (0.84,1.64) -- (0.64,1.44) -- (0.64,0) --cycle;
		\draw[thick,<->,>=latex] (0.29,1.19) --++ (.7,.7);

		\draw[ultra thick] (-2,1.64) to [out=0,in=270] (-1,3) -- (-1,3.5);
		\draw[thick,->,>=latex] (-2,2) --++ (.7,0);
		\draw[thick,->,>=latex] (-2,2.5) --++ (.95,0);
		\draw[thick,->,>=latex] (-2,3) --++ (1,0);
		\draw[thick,->,>=latex] (-2,3.5) --++ (1,0);
		\draw[thick] (-0.8,1.64) --++ (0,0.68) node [right] {$\delta$} --++ (0,0.68) ++ (-0.1,0) --++ (0.2,0);
		\draw[thick] (-2.8,0) ++ (-0.1,0) --++ (0.2,0) ++ (-0.1,0) --++ (0,0.82) node [left] {$H$} --++ (0,0.82) ++ (-0.1,0) --++ (0.2,0);
		\draw[thick,->,>=latex] (-3.5,2.5) --++ (.5,0) node [below] {$x$} --++ (.5,0);
		\draw[thick,->,>=latex] (-3.5,2.5) --++ (0,.5) node [left] {$y$} --++ (0,.5);
	\end{tikzpicture}
	\caption{Schematic of the actuator and an example of the phase-locked horizontal velocity field ($U_x/U_\infty$) for $St_H = 0.21$.
	\label{fig:SJA_schem}}
\end{figure}
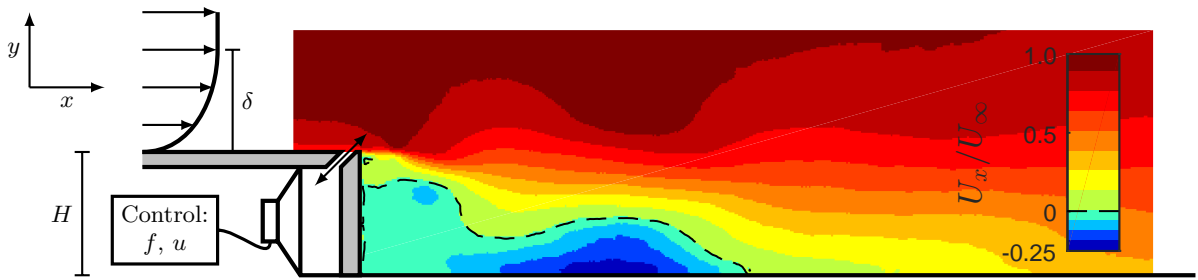

\subsection{Description of the step and separating flow}
The separating flow over a two-dimensional backward-facing step of height $H = 0.05$~m, spanning the entire width of the test section, is considered.
The incoming flow consists of a fully-developed turbulent boundary layer with a free-stream velocity of $U_\infty = 12$~m/s ($Re_H = HU_\infty/\nu = 41000$), a boundary layer thickness of $\delta_{99} = 0.074$~m and a friction Reynolds number of $Re_\tau \approx 2600$.
In absence of actuation, the flow separates from the top of the step and reattaches at a downstream location of $x_{R,0} = 0.226$~m ($x_{R,0}/H = 4.52$), measured from the base of the step.

\subsection{Description of the synthetic jet actuator}
The separating flow is forced using a synthetic jet at the edge of the step.
The synthetic jet is actuated using a series of in-phase Visaton SC 8 N speakers placed side-by-side.
The speakers are attached to a single cavity that has a neck with a length of 6~mm and placed at an angle of 45 degrees to both the horizontal and vertical face of the step. The jet is issued from a slit that has a width $d = 0.5$~mm.
Four forcing cases are considered with actuation frequencies $f = 50$, 116, 204 and 476~Hz, corresponding to Strouhal numbers of $St_H = 0.21$, 0.48, 0.85 and 1.98.
A sinusoidal velocity cycle as given in Equation~\ref{eq:ut} is used with a maximum velocity of $u=20$~m/s for all cases.
A schematic of the cross-section of the step and actuator is presented in Figure~\ref{fig:SJA_schem}.

\subsection{Measurement procedures}
Measurements are performed in the University of Southampton's suction wind tunnel.
The test section of this tunnel measures 4.5~m in streamwise direction, 0.9~m in spanwise direction and 0.6~m in wall-normal direction.
Velocity fields are determined using two-dimensional, two-component particle image velocimetry (PIV).
The PIV system consists of two aligned Litron 200~mJ dual-pulse Nd-YAG lasers and three LaVision Imager Pro LX 16MP cameras, fitted with 105~mm focal-length lenses.
The width of the laser sheet is approximately 1~mm.
Seeding is provided by a Martin Magnum~1200 smoke machine, producing particles with a mean diameter of 1~$\mu$m.
For each case, 480 image pairs per phase are recorded for eight equidistant phases, phase-locked to the actuation signal.
Vectors are determined using a first pass with a window size of 64~$\times$~64 pixels, followed by a second pass of 24~$\times$~24 pixels with 50\% overlap, resulting in a resolution of one vector per 0.7~mm.
Velocity fields are corrected for pixel-locking as described by~\citet{Hearst2015}.

The field of view extends between $-0.5<x/H<14$ in streamwise direction and $0<y/H<2$ in wall-normal direction, both measured with the origin located at the base of the step.
Since (with exception of the $St_H = 0.21$ case) vortices can only be tracked up to $x/H<6$ and all recirculation regions end before this value, all flow fields will be presented with streamwise extend $-0.5<x/H<6$.
The measurement plane is located at the spanwise centre of the wind tunnel and synthetic jet.
It is assumed that the flow in the measurement plane is largely two-dimensional.
The analysis is based on the measurement of the reattachment length and spanwise vortices, both of which are defined for in-plane components only and would not be influenced by unexpected out-of-plane motions.

\section{Results and discussion}
Time-averaged and phase-locked velocity and vorticity fields are presented for selected cases in Sections~\ref{sec:TA} and \ref{sec:PL} respectively.
The formation of the observed vortical structures is exemplified using phase-locked images in Section~\ref{sec:Train}.
Section~\ref{sec:Entrain} discusses the effect of these structures on the time-averaged entrainment and circulation.
Variations are explained by tracking vortex trajectories and magnitudes in Section~\ref{sec:Track}.
Relevant scaling parameters and equations are derived based on these observations and applied in Section~\ref{sec:Scaling}.
The discussion is concluded with the identification of different regimes of Strouhal number in Section~\ref{sec:Regime}.

\subsection{Time-averaged velocity and reattachment length}\label{sec:TA}
\begin{figure}
\includegraphics{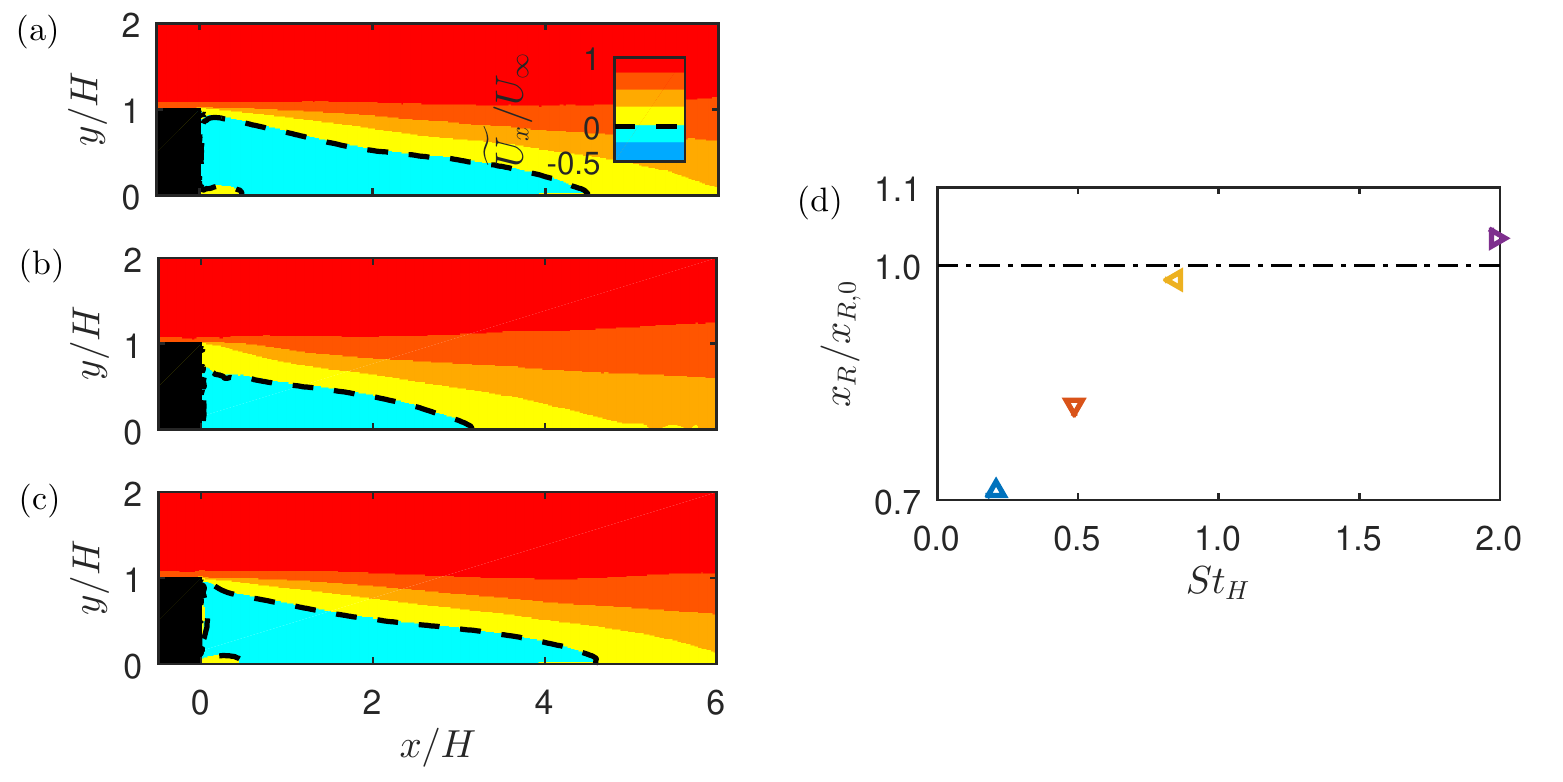}
\caption[Recirculation regions]{
	Time-averaged horizontal velocity component ($\widetilde{U_x}/U_\infty$), indicating the recirculation regions in absence of actuation (a), for $St_H = 0.21$ (b) and $St_H = 1.98$ (c).
	The dashed line indicates the contour for $\widetilde{U_x} = 0$.
	Reattachment lengths are compared in (d) where the dash-dotted line indicates the reattachment length in absence of actuation.
	\label{fig:Recirculation}
}
\end{figure}

Normalized time-averaged velocity fields showing the streamwise velocity component, $\widetilde{U}_x$, are presented in Figure~\ref{fig:Recirculation} for the baseline case without actuation (a) as well as for the $St_H = 0.21$ (b) and $St_H = 1.98$ (c) cases.
All cases show a clear recirculation region -- defined as the part of the flow where $\widetilde{U}_x < 0$ -- bounded by the dashed lines in Figure~\ref{fig:Recirculation}.
The reattachment length, $x_R$, is defined as the streamwise extent of the recirculation region for $y = 0$, i.e. the location at which the dashed line reaches the $x$-axis.
Reattachment lengths are quantified for all four Strouhal numbers and plotted in Figure~\ref{fig:Recirculation}d.
For low Strouhal numbers, the reattachment length is significantly shorter than in absence of actuation, indicated by the dash-dotted line.
The reattachment length increases with Strouhal number and grows beyond the unperturbed length for high Strouhal numbers.
Values of the normalized reattachment length, $x_R/H$, are tabulated for the unforced case and the four forcing cases in Table~\ref{tab:xr}.

\begin{table}
\caption{Reattachment length for all cases \label{tab:xr}}
\begin{ruledtabular}
\begin{tabular}{l c c c c c}
	$St_H$				& 0 	& 0.21	&  	0.48	& 0.85	&	1.98\\
	$f$~(Hz)			& 0 	&	50	&	116		& 204	&	476\\
	$x_R/H$				& 4.52 	& 3.22	&	3.71	& 4.44	& 	4.68\\
\end{tabular}
\end{ruledtabular}
\end{table}

As discussed in Section~\ref{sec:Intro}, the observed variations in reattachment length imply that there is a balance between increasing and decreasing effects of the forcing on this length.
Depending on the Strouhal number, this balance will lead to an elongated or a shortened reattachment length.
Following the literature, it is assumed that the main effect increasing the reattachment length is a stabilization of the shear layer by a reduction of disturbances in the incoming flow.
The main effect shortening the reattachment length is assumed to be entrainment of horizontal momentum into the recirculation region.

The alternating nature of the flow in the wall-normal direction as presented schematically in Figure~\ref{fig:St_Xr_schem}b is not present in the time-averaged velocity fields.
Phase-locked data is used to study these alternating components of the flow.

\subsection{Phase-locked velocity components and vorticity}\label{sec:PL}
\begin{figure}
\includegraphics{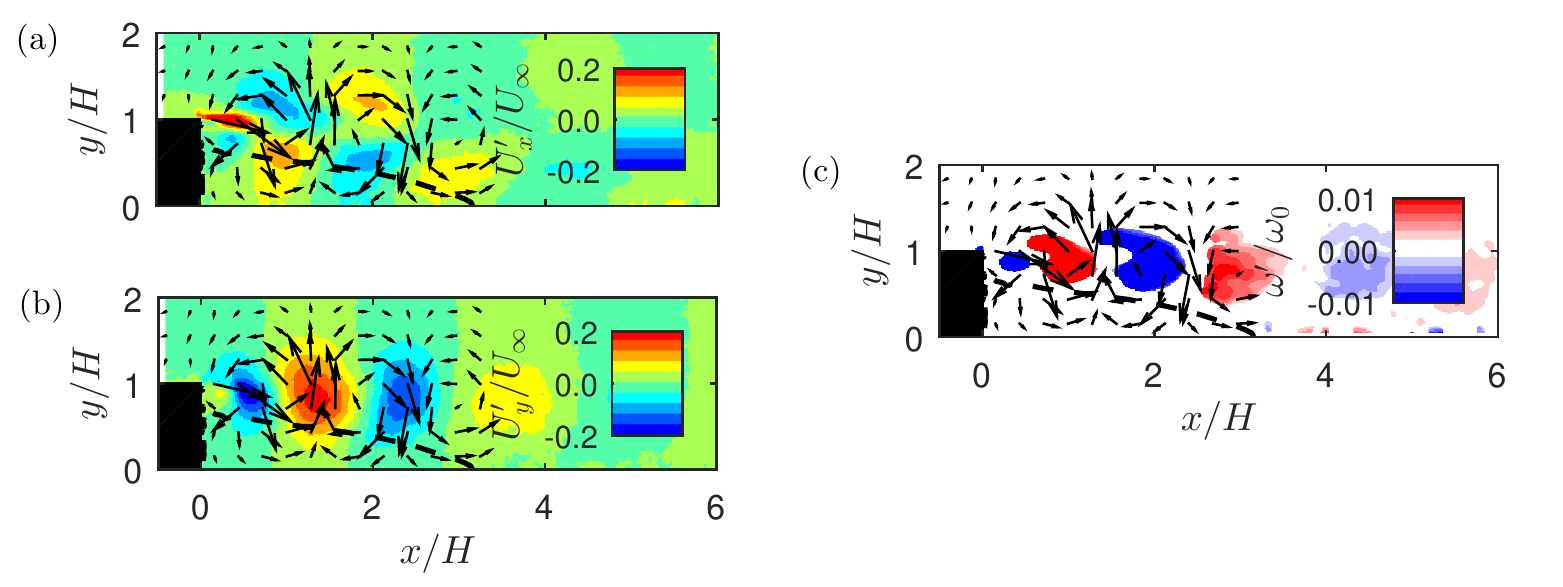}
\caption[Normalized phase-locked velocity and vorticity fields]{
	Normalized phase-locked horizontal- ($U'_x/U_\infty$, a) and vertical- ($U'_y/U_\infty$, b) velocity fluctuations for the $St_H = 0.21$ case.
	The vorticity ($\omega'/\omega_0$, c) is plotted at regions identified as vortices using swirling strength.
	\label{fig:Flow}
}
\end{figure}

Normalized phase-locked velocity and vorticity fields for a single phase of the $St_H = 0.21$ case are presented in Figure~\ref{fig:Flow}.
Observed structures are qualitatively similar for all forcing cases.
The flow fields are represented as fluctuations (normalised by $U_\infty$), which are obtained by subtracting the time-averaged velocity from the phase-averaged velocity ($U' = U_{phase} - \widetilde{U}$, where the tilde denotes a time-averaged quantity). Velocity vectors are overlaid on the velocity and vorticity maps.
Note that for clarity only one in 20 vectors is displayed. The normalized horizontal- and vertical velocity fluctuations, phase-locked near the end of the blowing phase, are presented in (a) and (b) respectively.

The horizontal velocity component shows a small jet at the top of the step, which is created by the blowing phase of the synthetic jet.
The direction of this jet shows that although the synthetic jet is injected into the flow at an angle of 45 degrees, the relatively high momentum of the cross-flow immediately deflects the jet into the horizontal direction.
Further downstream of the step an oscillating pattern of increased and decreased streamwise velocity components is visible.

The vertical velocity component (b) also shows an oscillating motion of positive and negative wall-normal velocities.
Positive wall-normal velocity transports low-momentum (or momentum of opposite direction) fluid out of the recirculation region, while negative wall-normal velocity injects high-momentum fluid into the recirculation region.
Both effects lead to an increase (or entrainment) of streamwise momentum in the recirculation region.

The normalized vorticity field corresponding to these velocity fields is presented in (c).
In this vorticity field the vorticity outside vortices (identified using swirling strength, see \citet{Adrian2000}) is set to zero.
The vorticity is calculated from the phase-averaged velocity fluctuations ($\omega' = \mathrm{d}U_y'/\mathrm{d}x - \mathrm{d}U_x'/\mathrm{d}y$) and normalized by $\omega_0 = \Gamma_j/A$, where $\Gamma_j$ is the circulation as given by Equation~\ref{eq:Gamma} and $A$ (m$^2$) is the volume of fluid exerted per velocity cycle, given by

\begin{equation}
A = L d = \int_0^T u_j(t) \mathrm{d}t \: d = \frac{u d}{\pi f},\label{eq:A}
\end{equation}

\noindent where $u_j(t)$ (s) is defined by Equation~\ref{eq:ut}, $L = u/(\pi f)$ (m) is the length of the theoretical slug of fluid ejected by the actuator each cycle and $d$ (m) is the slot width.
The vorticity field indicates that the alternating nature of the flow is caused by alternating clockwise and counterclockwise vortices.
Given that a synthetic jet creates pairs of counterrotating vortices (see for example \citet{Smith1998}), the pattern of single alternating vortices is surprising.
The formation of vortices is investigated by looking at the full vorticity field for multiple phases.

\subsection{Vortex formation and creation of train of vortices}\label{sec:Train}
To identify the mechanisms leading to the train of alternating vortices as observed in Figure~\ref{fig:Flow}c, phase-locked vorticity fields for the peak of the blowing cycle (a) and the peak of the suction cycle (b) are presented in Figure~\ref{fig:Creation}.
An annotated video showing all eight measured phases is available in the Supplementary Material.
Note that the vorticity fields in Figure~\ref{fig:Creation} show the full vorticity field including shear layers and vortices, in contrast to Figure~\ref{fig:Flow}c where the swirling strength is applied such that only vortices are displayed.
As shown in Figure~\ref{fig:Flow}a and discussed above, the jet is formed from the edge of the step directed in positive $x$-direction.
During the blowing phase (Figure~\ref{fig:Creation}a) this jet leads to a counterclockwise (red coloured) shear layer above it and a clockwise (blue coloured) shear layer beneath it.
In absence of a cross-flow, these shear layers would roll up into vortices, creating a vortex pair that convects away from the orifice and forms a train of vortex pairs \citep{Smith1998}.
However, during the suction phase (Figure~\ref{fig:Creation}b) two shear layers of opposite direction as during the blowing phase are formed due to interaction with the cross-flow.
Driven by the cross-flow, these shear layers convect in streamwise (positive-$x$) direction.
As can be seen from Figure~\ref{fig:Creation}, the positive shear layer created during the blowing phase connects to and merges with the positive shear layer created during the previous suction phase to form a single counterclockwise vortex.
In contrast, the negative shear layer created during the blowing phase merges with the negative shear layer created by the next suction phase, creating a single clockwise vortex.
The alternating forward and backward merging of the opposite vortices leads to the train of alternating vortices observed in Figure~\ref{fig:Flow}c and consequently to the alternating nature of the velocity fields as in Figures~\ref{fig:Flow}a~and~b.

\begin{figure}
\includegraphics[]{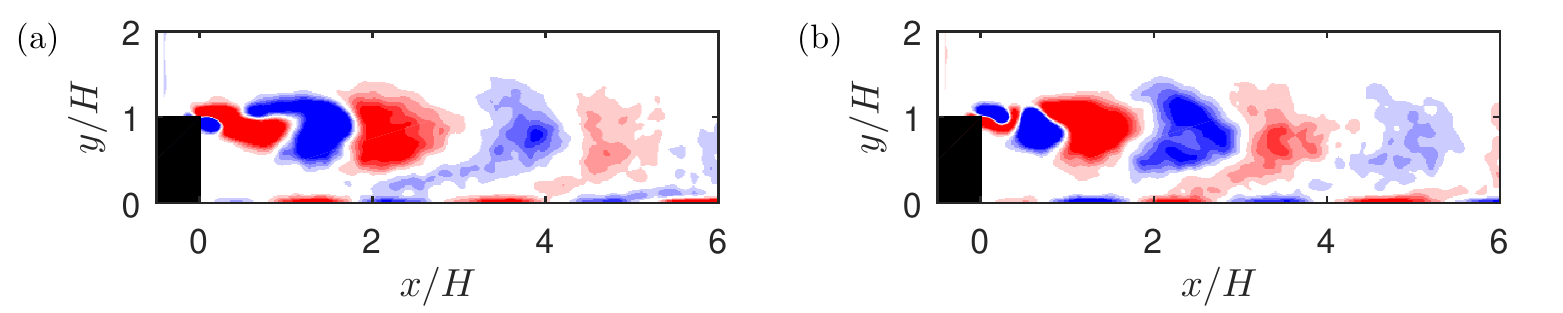}
\caption{Normalized phase-locked vorticity fields ($\omega'/\omega_0$) at the peak of the blowing cycle (a, $t/T=0.25$) and the peak of the suction cycle (b, $t/T=0.75$). See Figure~\ref{fig:Flow}c for colour scale.
	\label{fig:Creation}
}
\end{figure}

\subsection{Time-averaged entrainment}\label{sec:Entrain}
The described alternating train of vortices and corresponding alternating wall-normal velocity component as presented in Figure~\ref{fig:Flow}b leads to entrainment of momentum into the recirculation region. 
The downward motion injects high-momentum fluid in to the recirculation region while the upward motion removes low-momentum fluid from the recirculation region.
These two types of motion appear to be similar in nature in terms of vertical velocity. Therefore, to determine the entrainment, only the negative (downward) vertical component is conditionally averaged across all phases.
This serves as a proxy for entrainment and its variation is examined across different cases.

\begin{figure}
\includegraphics{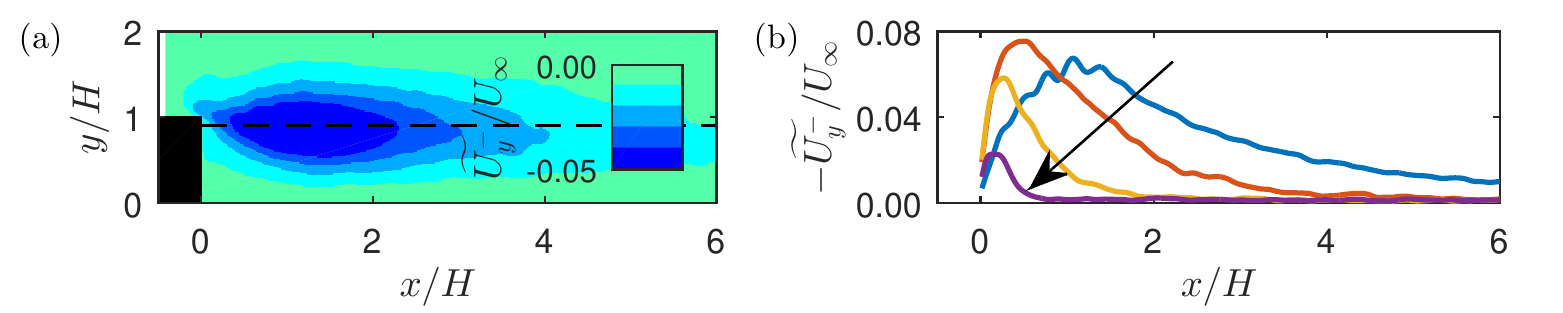}
\caption[Entrainment]{
	Time-averaged entrainment, defined as vertical velocity conditional averaged over negative velocities ($\widetilde{U_y^-}/U_\infty$) for the $St_H = 0.21$ case (a).
	The entrainment through the line $y/H=0.9$ for all cases as function of $x/H$ (b).
	$St_H=0.21$~({\color[rgb]{0,0.447,0.741} \rule[.4ex]{1em}{2pt}}), $St_H=0.48$~({\color[rgb]{0.850,0.325,0.098} \rule[.4ex]{1em}{2pt}}), $St_H = 0.85$~({\color[rgb]{0.929,0.694,0.125} \rule[.4ex]{1em}{2pt}}) and ${St_H=1.98}$~({\color[rgb]{0.494,0.184,0.556} \rule[.4ex]{1em}{2pt}}) (colour online), the arrow indicates increasing Strouhal number.
	\label{fig:entr_circ_1}
}
\end{figure}

This conditional time-averaged wall-normal velocity, $\widetilde{U_{y}^-} = \widetilde{U_y|}_{U_y<0}$ (meaning that the time average is taken over all instances of ${U_y<0}$), for the $St_H = 0.21$ case is presented in Figure~\ref{fig:entr_circ_1}a.
This velocity indicates the entrainment of volume per unit width and unit depth.
To compare different cases, the entrainment through the line $y/H = 0.9$, indicated by the dashed line, is considered.
The entrainment through this line represents the entrainment into the recirculation region and results are qualitatively independent of the exact vertical location of the line.
This entrainment as function of $x/H$ is presented in Figure~\ref{fig:entr_circ_1}b. 
The arrow indicates increasing Strouhal number.
The peak of maximum entrainment travels upstream for an increase in Strouhal number.
For the three lowest Strouhal number cases the peak entrainment is of similar magnitude but decreases more rapidly for the higher Strouhal numbers.
Maximum entrainment for the highest Strouhal number is significantly lower.

\begin{figure}
\includegraphics{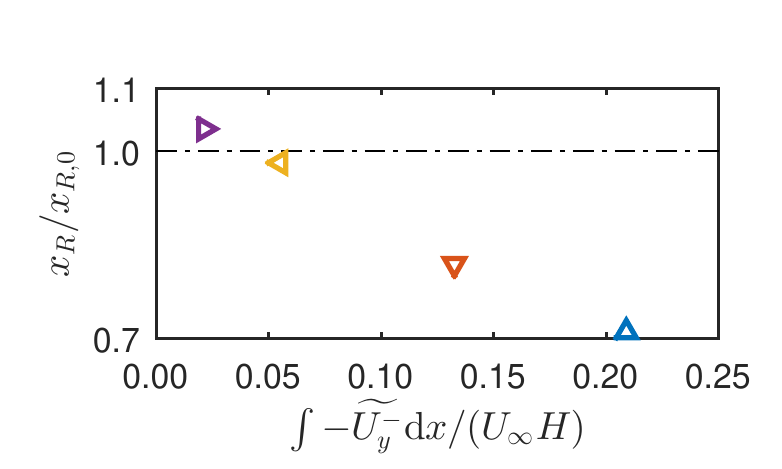}
\caption[]{Reattachment length as function of total entrainment for ${St_H=0.21}$~({\color[rgb]{0,0.447,0.741}\tiny $\bm{\mathrm{\Delta}}$}), $St_H=0.48$~({\color[rgb]{0.850,0.325,0.098}\tiny $\bm{\nabla}$}), $St_H = 0.85$~({\color[rgb]{0.929,0.694,0.125}$\bm{\triangleleft}$}) and ${St_H=1.98}$~({\color[rgb]{0.494,0.184,0.556}$\bm{\triangleright}$}).
	\label{fig:xr_entr}
}
\end{figure}

The total entrainment for each case can be calculated by integrating the curves in Figure~\ref{fig:entr_circ_1}b over streamwise distance, i.e. $\int -\widetilde{U_y^-} \mathrm{d}x$.
The relation between this total entrainment and the reattachment length is presented in Figure~\ref{fig:xr_entr}.
This graph shows a linear decrease in reattachment length with entrainment.
Comparing Figure~\ref{fig:xr_entr} with Figure~\ref{fig:Recirculation}d indicates that entrainment is a better predictor of reattachment length than the Strouhal number is, suggesting that entrainment is the driving force for changes in reattachment length.

\begin{figure}
\includegraphics{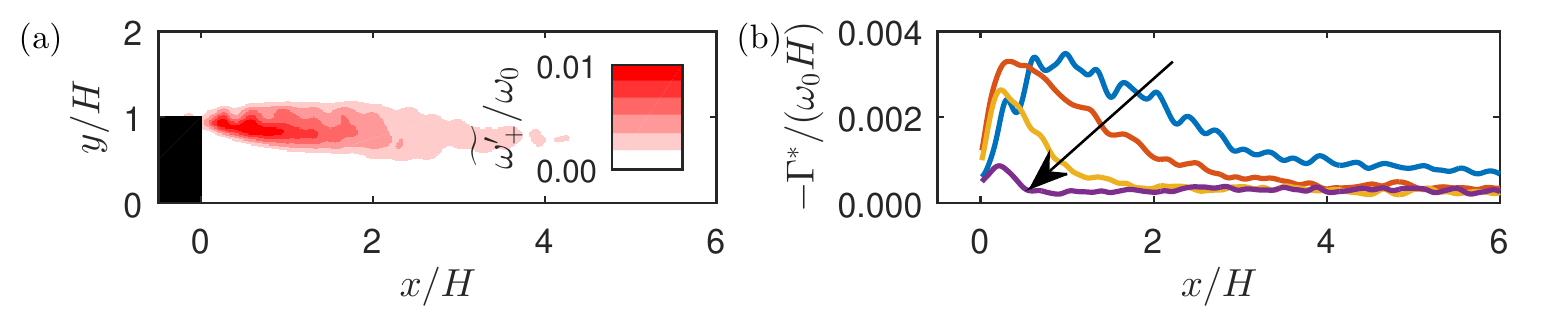}
\caption[Vorticity]{
	Time-averaged positive vorticity ($\widetilde{\omega'_+}/\omega_0$) with swirling strength applied for the $St_H = 0.21$ case (a).
	Time-averaged circulation per unit length ($\Gamma^*$) as function of $x/H$ for all cases (b).
	$St_H=0.21$~({\color[rgb]{0,0.447,0.741} \rule[.4ex]{1em}{2pt}}), $St_H=0.48$~({\color[rgb]{0.850,0.325,0.098} \rule[.4ex]{1em}{2pt}}), $St_H = 0.85$~({\color[rgb]{0.929,0.694,0.125} \rule[.4ex]{1em}{2pt}}) and ${St_H=1.98}$~({\color[rgb]{0.494,0.184,0.556} \rule[.4ex]{1em}{2pt}}) (colour online), the arrow in (b) indicates increasing Strouhal number.
	\label{fig:entr_circ_2}
}
\end{figure}

As discussed above, it is assumed that entrainment is primarily caused by vortices.
The time-averaged vorticity for the $St_H = 0.21$ case is presented in Figure~\ref{fig:entr_circ_2}a.
This time-average is conditioned to positive vorticity, $\widetilde{\omega'_+} = \widetilde{\omega'|}_{\omega'>0}$, with swirling strength applied, leading to the contribution of the counterclockwise vortices only.
Note that the contribution of clockwise vortices is equal but opposite as will be discussed below.
The vorticity distribution has roughly the same shape as the entrainment distribution presented in Figure~\ref{fig:entr_circ_1}a.
To compare the different cases, the vorticity is integrated over the wall-normal extent shown in Figure~\ref{fig:entr_circ_2}a.
This integral of vorticity can be seen as circulation per unit length in the streamwise direction and is given by

\begin{equation}
\Gamma^*(x) = - \int_0^{2H} \widetilde{\omega'_+}(x,y) \mathrm{d}y.
\end{equation}

This circulation is presented for all four cases in Figure~\ref{fig:entr_circ_2}b.
The arrow indicates increasing Strouhal number.
The shape of the circulation distribution is qualitatively similar to the entrainment distribution presented in Figure~\ref{fig:entr_circ_1}b for all cases and the entire range, indicating a strong link between circulation and entrainment.
For the three lowest Strouhal numbers the initial circulation reaches the same maximum, which is due to the strength and number of vortices balancing each other as discussed in the introduction.
The decrease in total entrainment for higher Strouhal numbers (and therewith the increase in reattachment length) can be attributed to a faster decay of circulation with streamwise distance for these Strouhal numbers.

To determine the mechanism behind this change in decay of circulation for higher Strouhal numbers, individual vortices are tracked, measuring their circulation and location as function of time.

\subsection{Vortex tracking}\label{sec:Track}
\begin{figure}
\includegraphics{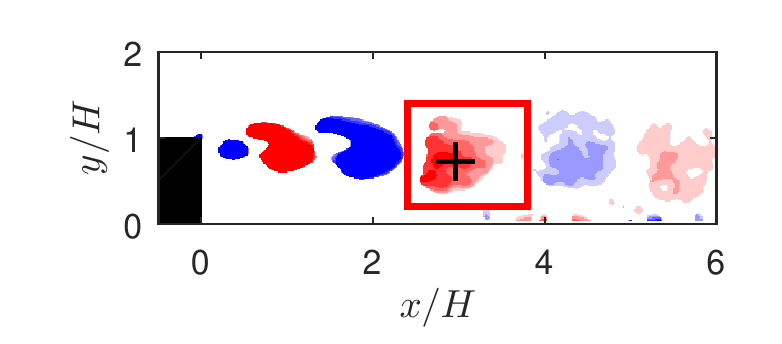}
\caption[Tracking method]{
	Example of the vortex tracking method. Vorticity is plotted at regions identified as vortices using swirling strength for the $St_H = 0.21$ case.
	See Figure~\ref{fig:Flow}c for colour bar.
	A vortex is identified inside the manually drawn rectangle.
	The calculated location of the vortex is identified by the black cross.
	\label{fig:Track}
}
\end{figure}

Vortices, as presented in Figure~\ref{fig:Flow}c for a single phase, can be tracked through time.
The tracking method is illustrated in Figure~\ref{fig:Track}.
Rectangles are manually drawn around the approximate region of a vortex.
Inside this region, vorticity of the appropriate sign is integrated to determine the circulation of the vortex, i.e.

\begin{equation}
\Gamma_\mathrm{CCW} = - \sum_{x_0}^{x_1} \sum_{y_0}^{y_1} \omega'(x,y)|_{\omega'>0} \Delta x \Delta y,
\end{equation}

\noindent where CCW depicts that we are looking at a counterclockwise (positive vorticity) vortex and $x_0$, $x_1$, $y_0$ and $y_1$ indicate the lower and upper bounds in horizontal and vertical direction respectively.
A similar equation is used for the clockwise vortices with the only difference being the condition that the vorticity is negative ($\omega'|_{\omega'<0}$).
The location of the vortex is defined as the weighted centre of vorticity, calculated as

\begin{equation}
x_\Gamma = \frac{1}{\Gamma} \sum_{x_0}^{x_1} \sum_{y_0}^{y_1} x \: \omega'(x,y) \Delta x \Delta y,
\end{equation}
\begin{equation}
y_\Gamma = \frac{1}{\Gamma} \sum_{x_0}^{x_1} \sum_{y_0}^{y_1} y \: \omega'(x,y) \Delta x \Delta y.
\end{equation}

For the vortex identified in Figure~\ref{fig:Track}, this leads to a circulation of $\Gamma = -0.11$ m$^2$s$^{-1}$ and a location of $x/H = 2.96$, $y/H = 0.73$.
The circulation and location are tracked throughout the field of view for all eight phases for the four cases.
Due to a decay in circulation of the vortices, at some point they can no longer be identified as coherent structures and tracking becomes impossible.
The limit for this tracking typically corresponds to the point where the circulation has decayed to about 10\% of the maximum circulation for each vortex.
This means that the faster decaying vortices at higher Strouhal numbers can be tracked for a shorter period of time.

\begin{figure}
\includegraphics[]{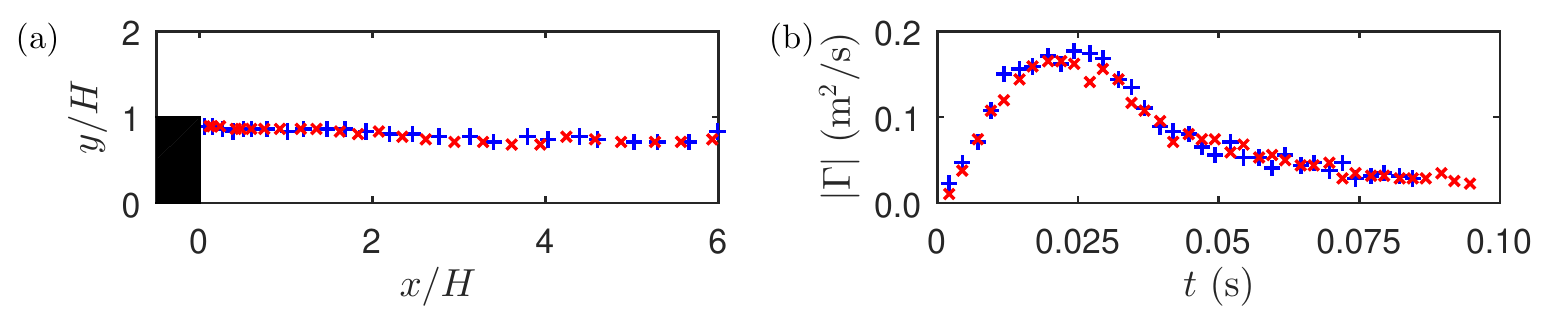}
\caption[Vortex tracking]{
	Comparison of tracked location (a) and absolute circulation (b) between positive~({\color[rgb]{0,0,1}\tiny $\bm{+}$}) and negative~({\color[rgb]{1,0,0}\tiny $\bm{\times}$}) vortex for the $St_H = 0.21$ case.
	\label{fig:Track_posneg}
}
\end{figure}

The tracked location (a) and circulation (b) for the positive and negative vortex independently for the $St_H = 0.21$ case are presented in Figure~\ref{fig:Track_posneg}.
Both the trajectories and the absolute values of circulation as function of time show a remarkable similarity between the positive (clockwise) and negative (counterclockwise) vortex.
The collapse of characteristics for the positive and negative vortex is observed for all four cases.
Because of this collapse the positive and negative vortices are treated as one in the subsequent analysis, effectively doubling the number of tracked points.

\subsubsection{Vortex trajectories}
\begin{figure}
\includegraphics[]{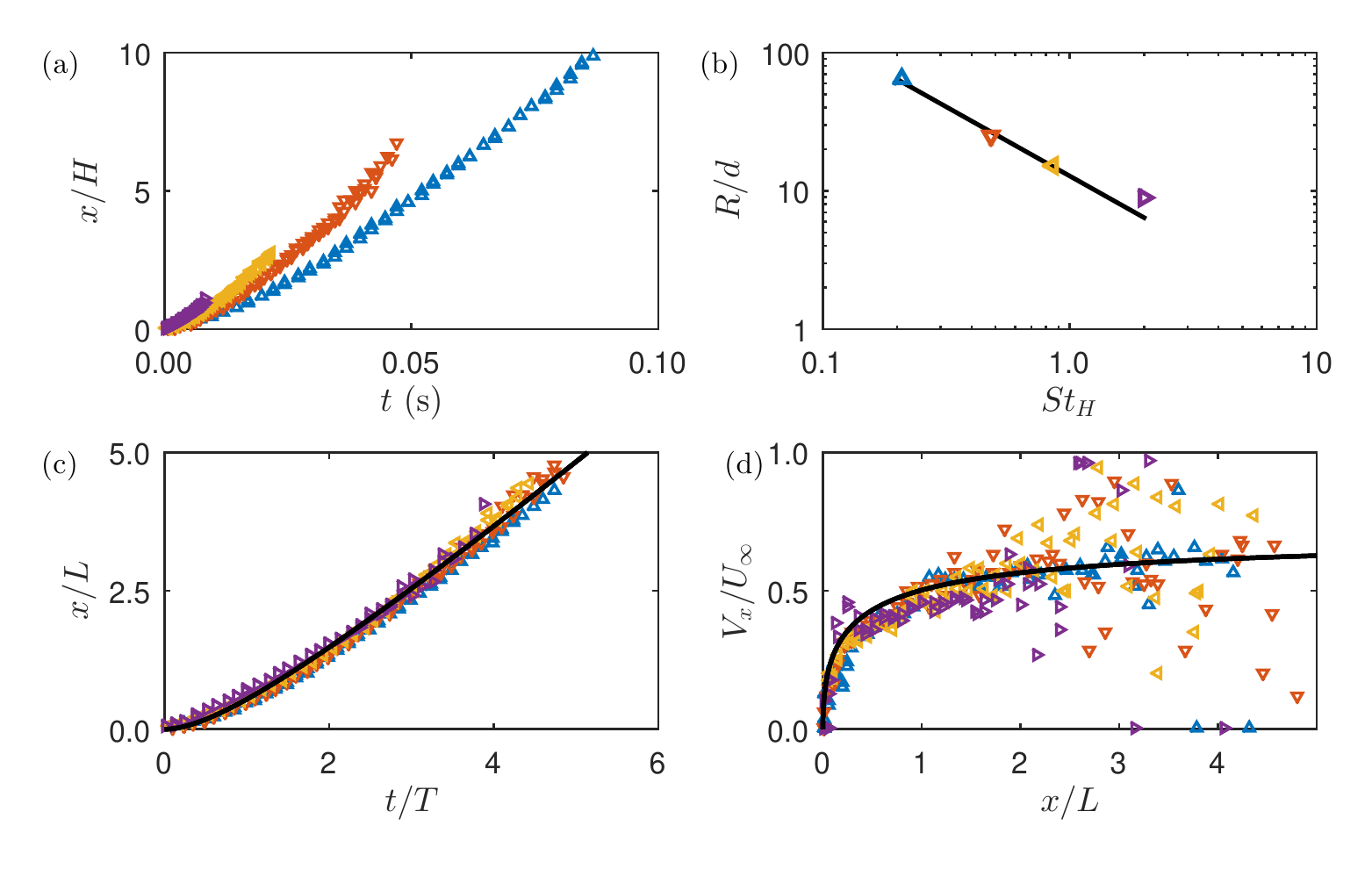}
\caption[]{
	Tracked vortex trajectories and size for ${St_H=0.21}$~({\color[rgb]{0,0.447,0.741}\tiny $\bm{\mathrm{\Delta}}$}), $St_H=0.48$~({\color[rgb]{0.850,0.325,0.098}\tiny $\bm{\nabla}$}), $St_H = 0.85$~({\color[rgb]{0.929,0.694,0.125}$\bm{\triangleleft}$}) and ${St_H=1.98}$~({\color[rgb]{0.494,0.184,0.556}$\bm{\triangleright}$}).
	Horizontal vortex location as function of time (a).
	Vortex radius as used for scaling (b) the black line indicates a decrease with 1/f.
	Scaled vortex location and time (c) the black line indicates the trajectory given by Equation~\ref{eq:x_t}.
	Normalized horizontal convection velocity of the vortices as function of scaled horizontal location (d) the black line indicates the convection velocity given by Equation~\ref{eq:V_t}.
	\label{fig:Trajectory}
}
\end{figure}

Vortex trajectories are tracked as presented for the ${St_H = 0.21}$ case in Figure~\ref{fig:Track_posneg}a.
For all cases, the vertical location is relatively constant and equal to the location of the jet.
The horizontal location as function of time is presented for all four cases in Figure~\ref{fig:Trajectory}a.
The vortices all start at the same location, $x=0$, with initial velocity $\mathrm{d}x/\mathrm{d}t = 0$.
Due to momentum transfer from the cross-flow to the vortices, they show an acceleration until they reach a relatively constant velocity.
Because of their smaller size, the acceleration is higher for the vortices corresponding to the high frequency case, which is manifested by the horizontal location increasing more rapidly with time in Figure~\ref{fig:Trajectory}a.

An analytical model for the trajectories can be derived by assuming that the (horizontal) acceleration of the vortices is driven by the local cross-flow velocity, $U_x^*$, impacting on half of the frontal area of the vortices, $R$ (with the other half being shielded by the step, see for example the first vortex in Figure~\ref{fig:St_Xr_schem}b).
The radii of the vortices for each case are presented in Figure~\ref{fig:Trajectory}b, where the black trend-line indicates a decrease of the radius as $R\propto1/f$.
This relation is unexpected given that the theoretical radius based on the (two-dimensional) volume $A$ given by Equation~\ref{eq:A} equals $R = \sqrt{A/\pi} = \sqrt{u d/(\pi^2 f)}$, scaling with $1/\sqrt{f}$.
The fact that the measured radius scales with $1/f$ can be explained by the vortices entraining fluid until a full train is formed, as is visible in Figures~\ref{fig:Flow} and~\ref{fig:Creation} where the alternating vortices are packed closely together.
Since the distance between vortices per definition scales with $1/f$, their radius must then scale in the same way.

Using this radius, the momentum transferred to each vortex by the local flow velocity can be described as

\begin{equation}
\frac{\mathrm{d}P}{\mathrm{d}t} = \rho R (U_x^* - V_x)^2,
\end{equation}

\noindent where $U_x^*$ is the local flow velocity and $V_x$ is the convection velocity of the vortex.
This momentum flux equals a force that will accelerate the vortex ($\mathrm{d}P/\mathrm{d}t = m \: \mathrm{d}V_x/\mathrm{d}t = \rho \pi R^2 \: \mathrm{d}V_x/\mathrm{d}t$), where the acceleration is then given by

\begin{equation}
\frac{\mathrm{d}V_x}{\mathrm{d}t} = \frac{(U_x^* - V_x)^2}{\pi R}.
\end{equation}

Assuming that $U_x^*$ is a constant (which is a reasonable assumption given that the wall-normal location of the vortices is constant as observed in Figure~\ref{fig:Track_posneg}) this can be integrated with respect to $t$ to give the convection velocity,

\begin{equation}
	\frac{V_x}{U_x^*} = \frac{U_x^* \: t/(\pi R)}{1+U_x^* \: t/(\pi R)}, \label{eq:V_t}
\end{equation}

\noindent where the initial condition $V_x(0) = 0$ is used.

Using the definition that $V_x = \mathrm{d}x/\mathrm{d}t$, this can be integrated with respect to $t$ again to give the trajectory,

\begin{equation}
	\frac{x}{\pi R} = U_x^* \frac{t}{\pi R} - \mathrm{ln}\left(1 + U_x^* \frac{t}{\pi R}\right), \label{eq:x_t}
\end{equation}

\noindent where the initial condition $x(0) = 0$ is used.

An important finding from Equations~\ref{eq:V_t}~and~\ref{eq:x_t} is that both $x$ and $t$ scale with $1/R$.
As discussed above, $R \propto 1/f$, implying that the period $T = 1/f$ or the slug length $L=u/(\pi f)$ can also be used for scaling.
The trajectories are replotted in Figure~\ref{fig:Trajectory}c, now with the location scaled as $x/L$ and the time scaled as $t/T$.
The normalized convection velocity of the vortices, with the location scaled as $x/L$, is presented in Figure~\ref{fig:Trajectory}d.
With the axes scaled in this way, the trajectories and convection velocities show a remarkable similarity across the four cases.
The value of $U_x^*$ is determined from the data as the convection velocity asymptotes to approximately $U_x^* = 0.7 U_\infty$.
The trajectory and velocity predicted by Equations~\ref{eq:V_t}~and~\ref{eq:x_t} is represented by the black lines in Figures~\ref{fig:Trajectory}c~and~d.
The model predicts the velocities and trajectories reasonably well for the complete data set.

\subsubsection{Vortex circulation}
\begin{figure}
\includegraphics[]{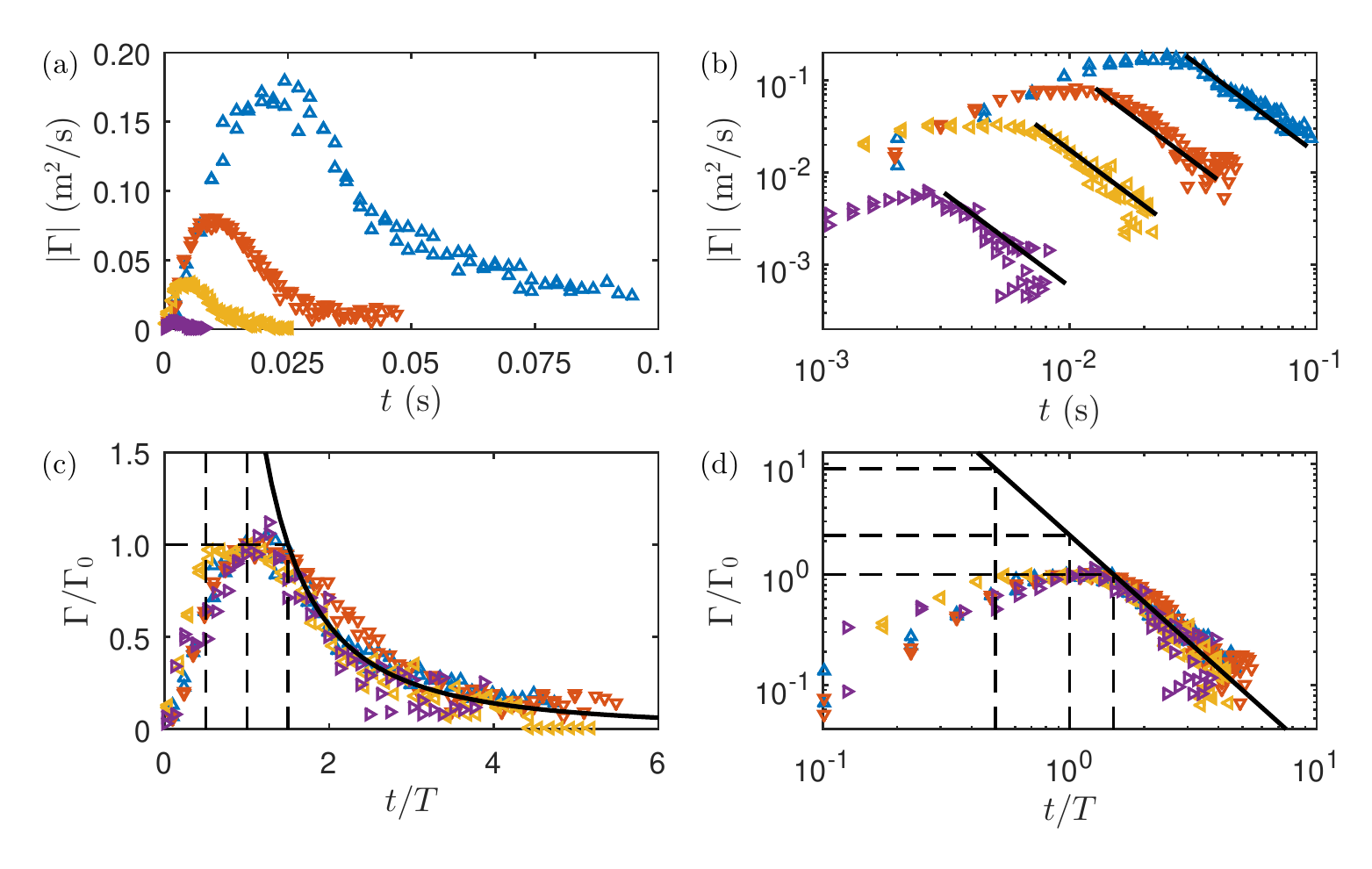}
\caption[]{
	(Absolute) circulation tracked as function of time for $St_H=0.21$~({\color[rgb]{0,0.447,0.741}\tiny $\bm{\mathrm{\Delta}}$}), $St_H=0.48$~({\color[rgb]{0.850,0.325,0.098}\tiny $\bm{\nabla}$}), $St_H = 0.85$~({\color[rgb]{0.929,0.694,0.125}$\bm{\triangleleft}$}) and ${St_H=1.98}$~({\color[rgb]{0.494,0.184,0.556}$\bm{\triangleright}$}).
	The data is plotted on (a) linear and (b) log-log axes.
	The black trend-lines in (b) show a decay of $t^{-2}$.
	Normalized circulation as function of scaled time on linear (c) and log-log axes (d).
	The black lines indicate the decay of circulation as given by Equation~\ref{eq:Gscaling}.
	Dashed lines indicate the values at $t/T = 0.5$, 1.0 and 1.5.
	\label{fig:Circulation}
}
\end{figure}

The absolute value of the circulation per vortex as function of time is presented in Figure~\ref{fig:Circulation}a for all four cases, all showing the same general trend.
Up to some point in time the circulation grows due to formation of the vortex.
As expected from Equation~\ref{eq:Gamma}, the magnitude of the peak circulation roughly scales with the inverse of frequency.
After this peak the circulation starts decaying.
The vortices can be tracked up to approximately 10\% of the peak circulation, which is longer in time for lower frequencies.

When plotted on log-log axes as in Figure~\ref{fig:Circulation}b, all four cases show the same decay rate.
The black lines overlaid on the data indicate a decay with $t^{-2}$, which corresponds well to the data.
This decay rate can be used to analyse the scaling between the cases.
An initial value of circulation ($\Gamma_0$) is reached at the start of the constant decay ($t_0$), i.e. $\Gamma(t_0) = \Gamma_0$.
This time $t_0$ is associated with the end of the formation of the vortex.
If we let $\Gamma(t) = C/t^2$ during the decay period, this constant $C$ can be determined using the initial condition, $C/t_0^2 = \Gamma_0$, leading to

\begin{equation}
\frac{\Gamma(t)}{\Gamma_0} = \left(\frac{t_0}{t}\right)^2,
\end{equation}

\noindent valid for the decay part only.
It can be determined empirically that the constant decay starts at $t_0 = 3/2 \: T$ for all four cases, leading to

\begin{equation}
\frac{\Gamma(t)}{\Gamma_0} = \frac{9}{4}\left(\frac{T}{t}\right)^2 = \frac{9}{4} \left(t f\right)^{-2}.\label{eq:Gscaling}
\end{equation}

This relation indicates that for adequate comparison between cases, time should be scaled with the actuation frequency.
The normalized circulation, $\Gamma(t)/\Gamma_0$, is plotted as function of the scaled time, $t/T$, in Figure~\ref{fig:Circulation}c~and~d.
As expected the data now collapses for the four cases.
This collapse is not only present during the decay period on which the scaling is based, but also during vortex formation.
The empirical model given by Equation~\ref{eq:Gscaling} is represented by the solid black lines in Figures~\ref{fig:Circulation}c~and~d.
Both on the linear and on the logarithmic axes, the data corresponds very well to this empirical model.

The reference circulation, $\Gamma_0$ (m$^2$/s), can be compared to the circulation theoretically produced by the synthetic jet, $\Gamma_j$, as given by Equation~\ref{eq:Gamma}.
The ratio $\Gamma_0/\Gamma_j$ for all four cases is presented in Table~\ref{tab:ratio}.
This ratio is approximately $\Gamma_0/\Gamma_j = 0.1$ for the three lowest Strouhal numbers, meaning that the measured circulation is significantly lower than the theoretically produced circulation.
This low ratio can be explained by assuming that as vortices are being formed, circulation is already decaying.
The empirical model for the decay is extrapolated in Figure~\ref{fig:Circulation}d to illustrate the influence of decay during the formation period.
Assuming a mean formation time for the circulation of $t=T/2$, i.e. half of the actuation cycle, Equation~\ref{eq:Gscaling} indicates a circulation of $\Gamma(T/2)/\Gamma_0 = 9$.
Using this new reference circulation, the ratio of measured versus expected circulation ($\Gamma(T/2)/\Gamma_j$) as presented in Table~\ref{tab:ratio} is much closer to one (for the lower three Strouhal numbers).

\begin{table}
\caption{Ratio of measured versus theoretically produced circulation \label{tab:ratio}}
\begin{ruledtabular}
\begin{tabular}{l c c c c}
	$St_H$				& 0.21	&  	0.48	& 0.85	&	1.98\\
	$f$~(Hz)			&	50	&	116		& 204	&	476\\
	$\Gamma_0/\Gamma_j$	& 0.10	&	0.10	& 0.08	& 	0.03\\
	$\Gamma(T/2)/\Gamma_j$	&	0.90	&	0.90	&	0.72	&	0.27\\
\end{tabular}
\end{ruledtabular}
\end{table}

For an increase in Strouhal number, most notably the highest Strouhal number of $St_H = 1.98$, Table~\ref{tab:ratio} shows a strong decrease in the ratio of expected versus theoretically produced circulation.
This decrease is likely caused by circulation cancellation due to close proximity of vortices to each other or ingestion of circulation due to close proximity of the vortices to the actuator when the suction phase starts.
The latter is analogue to the formation criterion as defined by \citet{Holman2005}, where for Strouhal numbers above the formation criterion no synthetic jet is formed due to ingestion of ejected fluid.

\subsection{Scaling parameters}\label{sec:Scaling}
\begin{figure}
\includegraphics[]{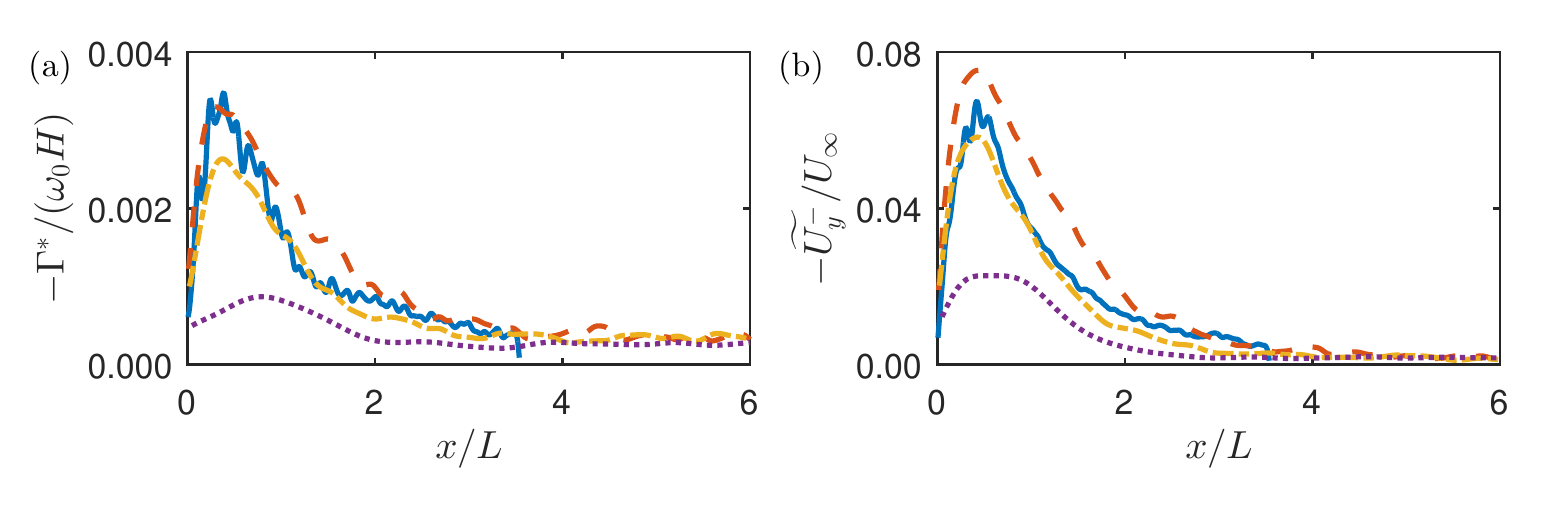}
\caption[Normalized vorticity and entrainment]{
	Time-averaged circulation per unit width (a) and entrainment through $y/H=0.9$ (b) as function of $x/L$.
	$St_H=0.21$~({\color[rgb]{0,0.447,0.741} \rule[.4ex]{1em}{2pt}}), $St_H=0.48$~({\color[rgb]{0.850,0.325,0.098} \rule[.4ex]{.35em}{2pt}\hspace{.2em}\rule[.4ex]{.35em}{2pt}}), $St_H = 0.85$~({\color[rgb]{0.929,0.694,0.125} \rule[.4ex]{.4em}{2pt}\hspace{.1em}\rule[.4ex]{.15em}{2pt}\hspace{.1em}\rule[.4ex]{.25em}{2pt}}) and ${St_H=1.98}$~({\color[rgb]{0.494,0.184,0.556} \rule[.4ex]{.2em}{2pt}\hspace{.2em}\rule[.4ex]{.2em}{2pt}\hspace{.2em}\rule[.4ex]{.2em}{2pt}}) (colour online).
	\label{fig:circ_entr_norm}
}
\end{figure}

With the time-evolution of both the vortex circulation and location known, the decay of circulation as presented in Figure~\ref{fig:entr_circ_2}b can be properly scaled.
Since both $\Gamma/\Gamma_0$ (see Figure~\ref{fig:Circulation}c) and $x/L$ (see Figure~\ref{fig:Trajectory}c) are functions of $t/T$ only, they can be rewritten to be functions of each other, i.e. $\Gamma/\Gamma_0 = f(x/L)$.
The total circulation measured in the flow field and the entrainment through the line $y/H = 0.9$ as presented in Figures~\ref{fig:entr_circ_1}b and \ref{fig:entr_circ_2}b are replotted with the horizontal axis normalized as $x/L$ in Figure~\ref{fig:circ_entr_norm}.
These graphs show that (apart from the highest frequency which has lower circulation as discussed above) the cases collapse well when the horizontal location is normalized as $x/L$.
It is the decay of the (normalized) circulation with frequency in Equation~\ref{eq:Gscaling} that causes the scaling of circulation with $x/L$ and therefore the decrease of entrainment for increasing frequencies.
The scaling does not only hold for the decay part on which it is based, but also for the formation part of the vortices.
This indicates that for lower Strouhal numbers the formation will be completed at a location further downstream, meaning that the created shear layers take a longer streamwise distance to roll up into vortices.
This implies that there is a lower limit on the Strouhal number, below which vortices will be formed at the end of (or beyond) the recirculation region and entrainment into the recirculation region will be minimal.
This leads to the effect of forcing on the reattachment length diminishing for low Strouhal numbers as is observed in the literature~\citep{Kiya1993,Chun1996}.

\subsection{Reconsideration of forcing regimes}\label{sec:Regime}
In the literature, frequency (Strouhal number) regimes of forcing are usually discussed in terms of low or high, where the low-frequency regime corresponds to shear-layer amplification whereas high-frequency corresponds to shear-layer stabilization.
Using combined insights from literature and the present study, the influence of the forcing frequency on the reattachment length could be explained differently.

\begin{figure}
	\begin{tikzpicture}
	\node [inner sep=0pt,above right] at (0,0)
	{\includegraphics{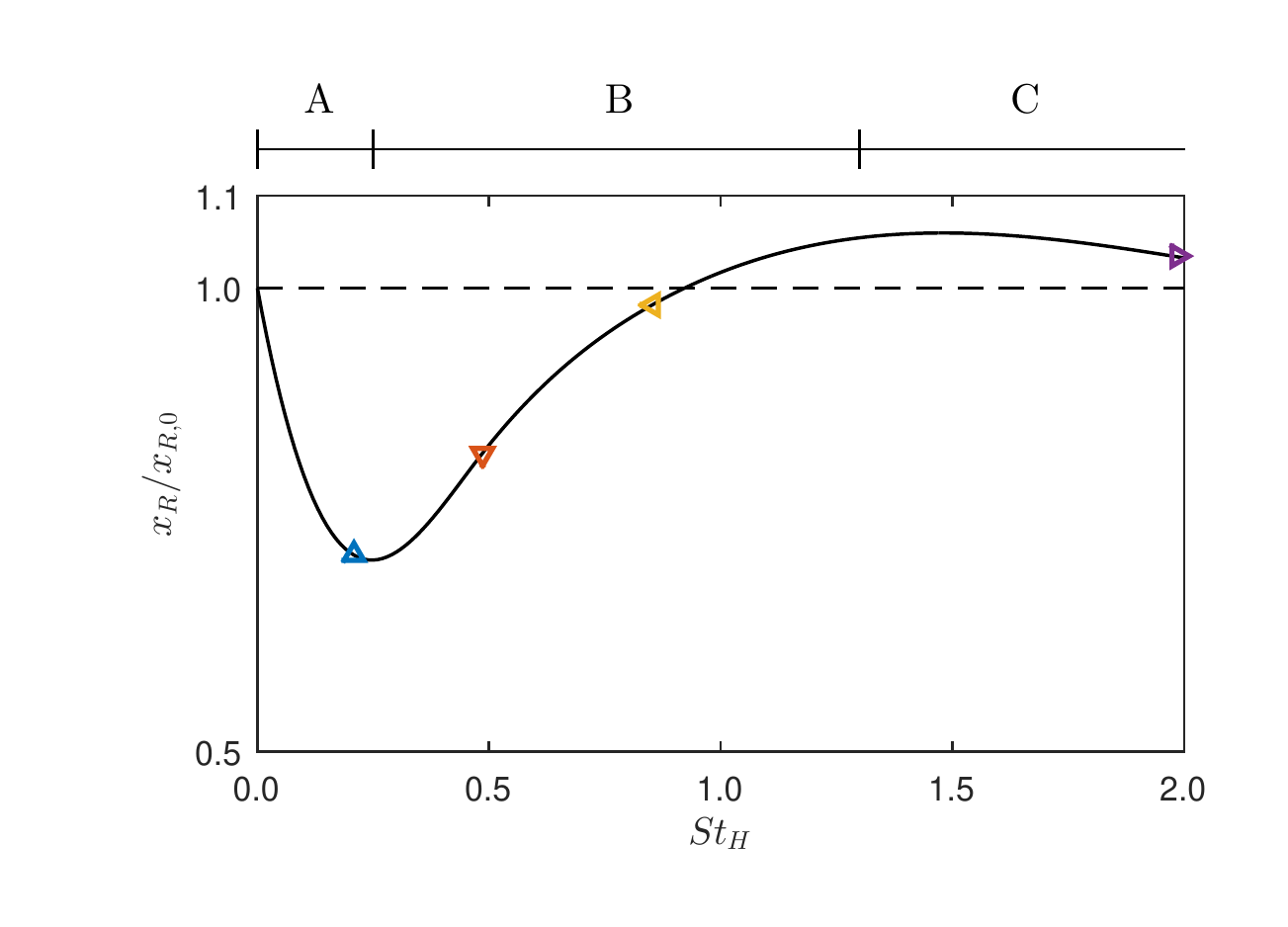}};
	
	\begin{scope}[xshift=4cm,yshift=3cm,scale=.6]
	\fill[lightgray] (-2,0) rectangle (0,-1);
	\draw[thick] (-2,0) -- (0,0) -- (0,-1) -- (5,-1);
	\draw[thick, dashed] (0,0) to [out=0,in=165] (4,-1);	
	
	\fill[red!50] (3.5,0) circle [radius = 0.7];
	\draw[line width = .12cm,red!50] (3.5,-.6) to [out=180,in=-30] (.4,.1) to ++(150:.1);
	\draw[thick] (3.5,0) circle [radius = 0.6];
	\draw[thick] (3.5,-0.6) to [out=180,in=-30] (.4,.1);
	\draw[thick,->] (4.1,0) --++ (0,.05);
	
	\fill[blue!50] (2,0) circle [radius = 0.65];
	\draw [line width=.12cm,blue!50] (2,.55) to [out=180,in=30] (.3,-.1) to ++(210:.1);
	\draw[thick] (2,0) circle [radius = 0.55];
	\draw[thick] (2,0.55) to [out=180,in=30] (.3,-.1);
	\draw[thick,->] (2.55,0) --++ (0,-.05);
	\end{scope}
	\draw[thick] (3.25,3.25) --++ (-.25,1.25);
	\draw[ultra thick,->,>=latex] (3.5,4.5) --++ (-.225,0.75) node [above right] {\large a} --++ (-.225,0.75);
	
	\begin{scope}[xshift=6.5cm,yshift =4.25cm,scale=.6]
	\fill[lightgray] (-2,0) rectangle (0,-1);
	\draw[thick] (-2,0) -- (0,0) -- (0,-1) -- (5,-1);
	\draw[thick, dashed] (0,0) to [out=0,in=165] (4,-1);
	
	\fill[red!50] (4.5,0) circle [radius = 0.1];
	\draw[thick] (4.5,0) circle [radius = 0.05];
	\draw[thick,->] (4.55,0) --++ (0,.05);
	
	\fill[blue!50] (3.7,0) circle [radius = 0.12];
	\draw[thick] (3.7,0) circle [radius = 0.07];
	\draw[thick,->] (3.77,0) --++ (0,-.05);
	
	\fill[red!50] (2.9,0) circle [radius = 0.16];
	\draw[thick] (2.9,0) circle [radius = 0.11];
	\draw[thick,->] (3.01,0) --++ (0,.05);
	
	\fill[blue!50] (2.1,0) circle [radius = 0.21];
	\draw[thick] (2.1,0) circle [radius = 0.16];
	\draw[thick,->] (2.26,0) --++ (0,-.05);
	
	\fill[red!50] (1.3,0) circle [radius = 0.35];
	\draw[thick] (1.3,0) circle [radius = 0.3];
	\draw[thick,->] (1.6,0) --++ (0,.05);
	
	\fill[blue!50] (.5,0) circle [radius = 0.5];
	\draw[thick] (.5,0) circle [radius = 0.45];
	\draw[thick,->] (.95,0) --++ (0,-.05);
	\end{scope}
	\draw[thick] (6,4.5) --++ (-.25,.85);
	\draw[ultra thick,->,>=latex] (4.25,4.75) --++ (0.75,0.75) node [above left] {\large b} --++ (0.75,0.75);
	
	\begin{scope}[xshift=9cm,yshift=5.5cm,scale=.6]
	\fill[lightgray] (-2,0) rectangle (0,-1);
	\draw[thick] (-2,0) -- (0,0) -- (0,-1) -- (5,-1);
	\draw[thick, dashed] (0,0) to [out=0,in=165] (4,-1);
	
	\fill[red!50] (1.14,0) circle [radius = 0.08];
	\draw[thick] (1.14,0) circle [radius = 0.03];
	\draw[thick,->] (1.17,0) --++ (0,.05);
	
	\fill[blue!50] (.96,0) circle [radius = 0.10];
	\draw[thick] (.96,0) circle [radius = 0.05];
	\draw[thick,->] (1.01,0) --++ (0,-.05);
	
	\fill[red!50] (.68,0) circle [radius = 0.18];
	\draw[thick] (.68,0) circle [radius = 0.13];
	\draw[thick,->] (.81,0) --++ (0,.05);
	
	\fill[blue!50] (.25,0) circle [radius = 0.25];
	\draw[thick] (.25,0) circle [radius = 0.20];
	\draw[thick,->] (.45,0) --++ (0,-.05);
	\end{scope}
	\draw[thick] (9.5,5.75) --++ (.5,1);
	\draw[ultra thick,->,>=latex] (9.5,7) to [out=-20,in=185, looseness=.5] ++ (2.25,-.3) --++ (.1,0);
	\draw (10.6,6.6) node [below] {\large c};
	
	\end{tikzpicture}
	\caption{Reattachment length as function of Strouhal number.
		The black line indicates the trend found in literature.
		Symbols represent cases from the present study (note that the trend-line is adapted to these cases).
		Regimes A, B and C are defined in the text.
		\label{fig:Regimes}}
\end{figure}
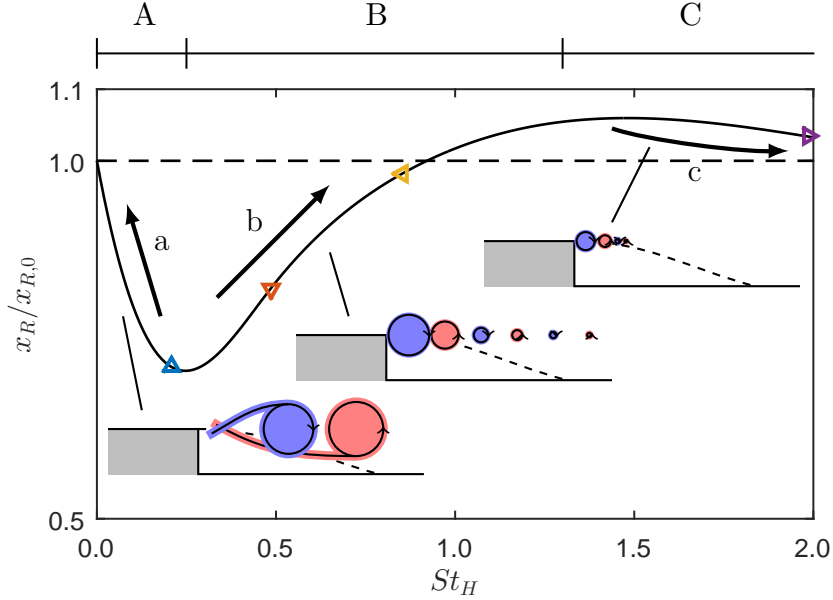

The reattachment length is presented as function of Strouhal number in Figure~\ref{fig:Regimes}.
The symbols correspond to the cases in the present study and are overlaid on the trend-line from literature discussed in Section~\ref{sec:Intro}.
Note that this trend is adapted to fit the data.
Using the results from literature and the present study, three regimes can be defined in this graph.
Schematic representations of vortical structures are embedded for regimes A, B and C.


In regime A, corresponding to low Strouhal numbers, the effect on the reattachment length decreases for decreasing Strouhal numbers.
In this regime the roll-up of vortices, scaling with $1/f$ takes a relatively long time and is completed at a point far downstream compared to the recirculation region.
As discussed above, this will lead to less entrainment into the recirculation region.
This is visible in Figure~\ref{fig:Regimes} by the effect on the recirculation region decreasing for a decreasing Strouhal number in regime A (in the direction of arrow a).

In regime B, for moderate Strouhal numbers, a train of vortices is formed.
The circulation, entrainment and reattachment length are dominated by the decay of circulation with $\Gamma(t) \propto (tf)^{-2}$ as given by Equation~\ref{eq:Gscaling} and discussed above.
This results in the circulation decaying more rapidly for higher frequencies, resulting in lower entrainment and an increase in reattachment length with Strouhal number as visible in regime B in Figure~\ref{fig:Regimes} (in the direction of arrow b).
The increasing reattachment length has the ability to overshoot the unforced length.
It is shown by~\citet{Dandois2007} that this is caused by the synthetic jet decreasing the amplitude of incoming perturbations, thereby lowering entrainment and increasing the reattachment length.

In regime C, corresponding to high Strouhal numbers, the created vortices are very close to each other and close to the actuator during the suction phase.
This leads to vorticity cancellation and re-ingestion into the actuator, meaning that the vortices contain relatively less circulation.
This implies that the strength of the synthetic jet decreases with Strouhal number in this regime, analogue to the formation criterion~\citep{Holman2005}.
The decrease in strength leads to a decrease in the effect on the reattachment length as is visible for increasing Strouhal numbers in regime C in Figure~\ref{fig:Regimes} (in the direction of arrow c).
Following the literature, it is assumed that the reattachment length stabilises at the unforced length ($x_R = x_{R,0}$) for high Strouhal numbers \citep{Kiya1993}.

\section{Conclusions}
The flow over a backward facing step is forced using a synthetic jet at the edge of the step.
The effect of the forcing Strouhal number on the reattachment length behind the step is studied experimentally.
Across all four studied cases, the reattachment length decreases linearly for increasing entrainment.
It is shown that entrainment is driven by vortices created by the forcing.
The formation and evolution of these vortices depends on the Strouhal number.
Three Strouhal number regimes have been identified.
For low Strouhal numbers, associated with large time and length scales, the formation of vortices occurs at a relatively far downstream location in the flow. This minimizes the interaction between formed vortices and the recirculation region.
For high Strouhal numbers, associated with small length scales, vortices of opposite orientation are spaced closely together, leading to cancellation of vorticity.
Furthermore, in this regime, vortices created by the synthetic jet have not travelled far enough downstream by the time the suction period of the jet starts, leading to ingestion of vorticity.
Both effects lead to a reduction of circulation and therewith a reduced effect of the forcing.
In the regime associated with intermediate Strouhal numbers a train of alternating vortices is formed that interacts with and entrains fluid into the recirculation region.
The location and circulation of vortices is tracked using phase-locked PIV data.
The convection velocity of the vortices can be described using an analytical model based on the impulse of the cross-flow accelerating the vortices.
The decay of vortex circulation in time is described by an empirical model showing a decay of circulation with $(tf)^{-2}$.
These models reveal the proper scaling factors ($x/L$ and $t/T$) for the trajectories and circulation.
Using these scaling factors, the data for all four cases collapses.
The derived scaling explains why the total circulation decreases for an increasing Strouhal number, leading to a decrease in entrainment and elongation of the reattachment length.
Because of a decrease in entrainment for high Strouhal numbers, combined with an amplitude reduction of incoming perturbations, forcing has the ability to increase the reattachment length compared to the unforced case.

\bibliography{PRF_BERK}

\end{document}